\documentclass[journal=apchd5, manuscript=article]{achemso}

\usepackage[T1]{fontenc}       
\usepackage[hidelinks]{hyperref}
\usepackage{graphicx}
\usepackage{amsmath}
\usepackage{amssymb}
\usepackage{bbold}
\usepackage{xcolor}
\usepackage{bm}
\usepackage[normalem]{ulem} 
\usepackage{mathtools,accents}

\DeclareMathAlphabet\mathbfcal{OMS}{cmsy}{b}{n}

\SectionNumbersOff

\author{Mar\'ia Blanco de Paz}
\affiliation{Instituto de Qu\'imica F\'isica Blas Cabrera (IQF), CSIC, 28006 Madrid, Spain}
\author{Juan R. Deop-Ruano}
\affiliation{Instituto de Qu\'imica F\'isica Blas Cabrera (IQF), CSIC, 28006 Madrid, Spain}
\author{Diego M. Sol\'is}
\email{dmartinezsolis@com.uvigo.es}
\affiliation{Departamento de Teor\'ia de la Se\~nal y Comunicaciones, University of Vigo, 36301 Vigo, Spain}
\author{Alejandro Manjavacas}
\email{a.manjavacas@csic.es}
\affiliation{Instituto de Qu\'imica F\'isica Blas Cabrera (IQF), CSIC, 28006 Madrid, Spain}

\title{Lattice Resonances in Periodic Arrays of Time-Modulated Scatterers}


\begin{document}

\begin{abstract}
 
Lattice resonances are collective optical modes supported by periodic arrays of scatterers, arising from their coherent interaction enabled by the underlying periodicity. Owing to their collective nature, these resonances produce optical responses that are both stronger and spectrally narrower than those of individual scatterers. While such phenomena have been extensively studied in conventional time-invariant systems, recent advances in time-varying photonics present new opportunities to exploit and enhance the extraordinary characteristics of these collective modes. Here, we investigate lattice resonances in periodic arrays of time-modulated scatterers using a simple framework based on the dipolar approximation and time-Floquet theory, where each scatterer is modeled as a harmonic oscillator with periodically varying optical properties. We begin by analyzing the response of an individual scatterer, leveraging our model to identify the complex eigenfrequencies that define its dynamics. We show that, for the appropriate modulation amplitude and frequency, the imaginary part of one of these eigenfrequencies vanishes, leading to amplification. Building on this, we extend our analysis to a periodic array to investigate the effect of the interplay between temporal modulation and lattice resonances. In contrast to isolated scatterers, the collective nature of lattice resonances introduces a markedly more intricate spectral dependence of the amplification regime. Notably, this amplification emerges at substantially lower modulation strengths, facilitated by the enhanced light-matter interaction and increased lifetime provided by these collective resonances. Our work establishes a simple theoretical framework for understanding collective lattice resonances in time-modulated arrays,  enabling dynamic control and amplification of these modes. These findings open pathways toward next-generation nanoscale light sources and the design of active, nonreciprocal photonic devices.
\end{abstract}

\section{Introduction}

When multiple identical scatterers are arranged in a periodic array, their individual responses can couple through coherent multiple scattering, giving rise to collective optical modes known as lattice resonances \cite{WRV18,KKB18,UZR21}. These modes originate from the interplay between the intrinsic resonances of the scatterers and the diffractive coupling enabled by the periodicity of the array \cite{CBW19,ama91}. Although lattice resonances are fundamentally a wave phenomenon and therefore agnostic to the specific nature of the scatterers, they are most commonly studied in arrays of metallic nanostructures \cite{ZS04,ZJS04,AB08,HB14,WRV18,KKB18,UZR21}, where they stem from the coherent coupling of localized surface plasmon resonances, and in arrays of dielectric nanostructures\cite{ERS10,WKP19,CBG19,MCR20,ZXZ22,LTH22,ZZY23,ZLL23}, where they emerge from the coupling of Mie-like resonances. In both cases, the periodic arrangement produces optical responses that are significantly sharper\cite{ZRG17,LZG19,BRM21,GPS25} and stronger\cite{NKD12,ama68} than those of isolated elements, making lattice resonances a versatile platform for a variety of applications including sensing\cite{AYA09,DTW18,MHG18}, energy harvesting \cite{MCH23, ZBU14,ama91}, light-to-heat transduction\cite{BRK22,ama83}, and nanoscale light emission \cite{LLR13,ZDS13,SK14_2,RLV16,GNH19,HAH23}.

Lattice resonances have been extensively investigated in conventional structures composed of time-invariant scatterers, where the optical properties of the constituents remain constant in time. However, the recent advent of time-varying photonics, devoted to studying systems whose optical properties are structured in time at rates comparable to the operating frequencies, has unlocked new opportunities for dynamic control and functionalities that surpass the fundamental constraints of their static counterparts\cite{E23}. Temporal modulation enables the breaking of traditional constraints imposed by energy conservation, reciprocity, and passivity\cite{GTY22,CD19,CD19_2,WKH24}, granting access to regimes where light can be manipulated in unconventional ways, amplified, or even generated from vacuum fluctuations \cite{GHP19,KAD21,LLD22,GVL24,SGH25}. Time-varying photonic systems thus support a broad range of functionalities, including the realization of temporal boundaries\cite{XMA14,SKE21,MXY23,GMS25}, optical isolation\cite{YF09, CTD17,K23}, and nonreciprocal responses\cite{SKS15,SA17,CAT18}, among others. Moreover, periodic temporal modulations give rise to photonic time crystals, whose optical properties exhibit momentum-space band gaps within which electromagnetic waves can undergo exponential amplification \cite{RH15,LSS23,SSF23,AGW24}. Time-varying photonic systems are also able to mimic motion-induced effects without requiring actual mechanical movement, thereby enabling functionalities that would otherwise demand unattainable velocities \cite{HGG19,MA19_2,PS23,HVR25,amaXX}.

Several platforms have been proposed for implementing temporal modulation. At high frequencies, semiconductors and transparent conducting oxides enable strong modulation of the optical response in the IR and near-IR via ultrafast laser-induced changes in the carrier energy distribution \cite{ZLB16,VBV18,TVG23} and density \cite{SHP24}. In the GHz range, electro-optic modulation has been demonstrated in materials such as silicon\cite{XSP05,XMS07} and lithium niobate \cite{WZS18}. At the lower end of the spectrum, in the microwave regime, modulation is typically achieved electronically through tunable circuit elements that dynamically adjust capacitance or inductance \cite{ESS14,MLR21,PCL22}.

Most prior research on time-varying photonics has focused on extended systems such as surfaces and bulk materials. More recently, attention has shifted toward subwavelength scatterers \cite{MKP22,PLK23,VH25,GGH25}, whose optical resonances enhance light-matter interactions\cite{S18} facilitating effects like parametric amplification \cite{ALP22}. The combination of spatial periodicity with temporal modulation introduces additional degrees of freedom that enable new phenomena, including tilted band structures allowing for nonreciprocal propagation \cite{PAP23,WMA23,WGM25,GAZ25,GFL25}. 

Here, we present a comprehensive theoretical study of lattice resonances in periodic arrays of time-modulated scatterers. First, we develop a simple yet powerful framework that combines the dipolar approximation with time-Floquet theory, modeling each scatterer as a harmonic oscillator whose optical properties vary periodically. Using this approach, which fully accounts for dispersion, as well as radiative and nonradiative losses, we analyze the eigenmode structure of a time-modulated individual scatterer by tracking the evolution of its complex eigenfrequencies from the homogeneous (undriven) problem, as a function of the modulation frequency and amplitude. We identify the modulation conditions under which one of these eigenfrequencies crosses the real axis, signaling the emergence of an amplification regime. Second, exploiting a coupled dipole model, we extend our analysis to a periodic array of scatterers to explore how temporal modulation modifies the properties of lattice resonances and to investigate how the characteristic phenomena induced in time-varying photonic systems can be enhanced by the properties of these collective modes. Specifically, we calculate the absorbance of the periodic array and determine the modulation frequencies and amplitudes that give rise to amplification. Our results demonstrate a more intricate spectral response than that of an individual scatterer, attributable to the collective behavior of the lattice resonance, whose characteristics exhibit a strong dependence on the material and geometrical parameters of the array. Furthermore, the enhanced light-matter interaction and longer lifetimes provided by lattice resonances substantially decrease the modulation strength required to achieve amplification relative to the individual scatterers. This work sheds light on the critical role of collective effects in time-varying photonic systems and unveils new opportunities for dynamic amplification, active control, and nonreciprocal functionalities, offering an alternative route to achieve amplification without the need for a gain material. 

\section{Results and Discussion}

\begin{figure}[h!]
\begin{center}
\includegraphics[width=0.5\textwidth]{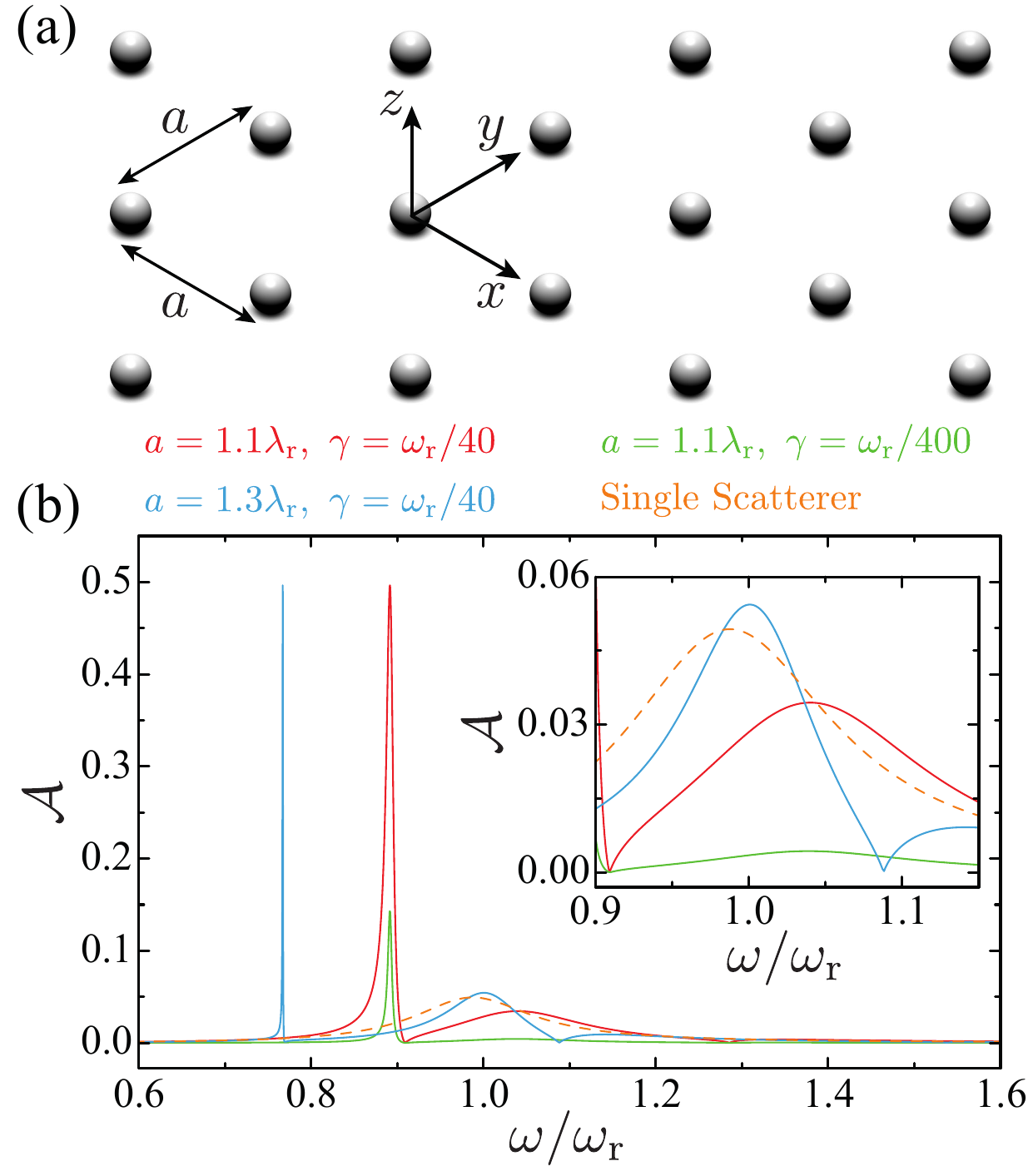}
\caption{(a) Schematic representation of the system under consideration, consisting of a square array of identical scatterers. The array lies in the $xy$ plane, is surrounded by vacuum, and is excited by a plane-wave electric field that propagates parallel to the $z$ axis and is polarized along the $x$ axis. (b) Absorbance spectra for time-invariant arrays with various combinations of period $a$ and nonradiative damping rate $\gamma$, as indicated in the legend. For reference, the orange dashed curve shows the corresponding results for an individual scatterer. The inset provides a zoomed view around the resonance of the individual scatterer. In all cases, we assume $\kappa = 0.15\omega_{\rm r}$.} \label{fig1}
\end{center}
\end{figure}

The system under consideration consists of a square array of isotropic scatterers with period $a$, as depicted in Figure~\ref{fig1}(a). We assume that the array is located in the $xy$ plane and surrounded by vacuum. Each of the scatterers is modeled as a point dipole, whose dipole moment satisfies the following equation of motion\cite{MK1976,ama37}
\begin{equation}
\mathbf{\ddot{p}}(t) + \omega_{\rm r}^2(t) \mathbf{p}(t) + \gamma \mathbf{\dot{p}}(t) - \frac{\kappa(t)}{\omega_{\rm r}^2(t)} \mathbf{\dddot{p}}(t)  = \frac{3c^3}{2}  \frac{\kappa(t)}{\omega_{\rm r}^2(t)}  \mathbf{E}(t), \label{eq1}
\end{equation}
where $c$ is the speed of light, $\mathbf{E}(t)$ is the external electric field, $\omega_{\rm r}$ denotes the resonance frequency of the individual scatterer, and $\gamma$ and $\kappa$ represent the nonradiative and radiative damping rates, respectively. The presence of the latter ensures that the optical response of the dipole satisfies the optical theorem\cite{MK1976,T03,ama98}, which is essential for accurately describing lattice resonances, as these are collective modes that arise from far-field coupling\cite{ama81}.

We consider the optical response of each individual scatterer to originate from the excitation of free carriers and introduce a periodic temporal modulation by varying their number. Under this general assumption, our framework is applicable to a wide variety of physical systems, including metallic structures operating at microwave frequencies, doped semiconductors in the IR and near-IR regions, and plasmonic platforms in the visible spectrum. The periodic modulation of the number of free carriers induces a corresponding variation in both the resonance frequency and the radiative damping rate, as the former scales with the square root of the number of carriers, whereas the latter scales with its square. We disregard any effect of this modulation on the nonradiative losses since those are expected to be much smaller. Accordingly, we have that $ \omega_{\rm r}^2(t) =  \omega_{\rm r}^2 f(t)$ and $\kappa(t)/\omega_{\rm r}^2(t) =  \tau f(t)$, where $\tau = \kappa/\omega_{\rm r}^2$ and $f(t) =  1+\Delta\cos(\Omega t)$, with $\Delta$ and $\Omega$ denoting the amplitude and frequency of the modulation, respectively.

To characterize the response of the scatterer, we calculate its polarizability, defined as
\begin{equation}
\mathbf{p}(t) = \int_{-\infty}^{\infty} {\rm d}t' \alpha(t,t') \mathbf{E}(t'). \nonumber
\end{equation}
which satisfies $\alpha(t,t')=0$ for $t'>t$ and fully accounts for the effect of dispersion\cite{SE21}. The periodic nature of the temporal modulation allows us to invoke Floquet theorem and expand both the induced dipole and the external field as 
\begin{equation}
\mathbf{p}(t) = \sum_{n=-\infty}^\infty \mathbf{p}_n {\rm e}^{-{\rm i}\omega_n t}, \nonumber
\end{equation} 
and 
\begin{equation}
\mathbf{E}(t) = \sum_{n=-\infty}^\infty \mathbf{E}_n {\rm e}^{-{\rm i}\omega_n t}, \nonumber
\end{equation} 
where $\omega_n=\omega + n\Omega$ and $\omega$ is the frequency of excitation. Substituting these expansions into Equation~\eqref{eq1}, we obtain
\begin{equation}
\mathbf{p}_n = \sum_{n'=-\infty}^\infty \alpha_{n,n'} \mathbf{E}_{n'}. \nonumber
\end{equation}
The coefficients $\alpha_{n,n'}$, which quantify how much the external field harmonic $n'$ polarizes the scatterer at harmonic $n$, are defined as (see the Supporting Information)
\begin{equation}
\alpha_{n,n'} = \int_{-\infty}^{\infty} {\rm d}t  \int_{-\infty}^{\infty} {\rm d}t' \alpha(t,t') {\rm e}^{{\rm i}\omega_n t}  {\rm e}^{-{\rm i}\omega_{n' }t'}, \nonumber
\end{equation}
and satisfy the relationship
\begin{equation}
A_n \alpha_{n,n'} + B_{n+1} \alpha_{n+1,n'} + B_{n-1} \alpha_{n-1,n'} = \frac{3c^3}{2}\tau\left[\delta_{n,n'} + \frac{\Delta}{2}\left(\delta_{n+1,n'} +\delta_{n-1,n'}\right)\right], \label{eq2}
\end{equation}
where $A_n = \omega_{\rm r}^2 - \omega_n^2-{\rm i}(\gamma \omega_n + \tau \omega_n^3)$ and $B_n = (\omega_{\rm r}^2 - {\rm i} \tau \omega_n^3)\Delta/2$. 
By truncating the expansions in $n$ and $n'$ at a sufficiently large value $N$, Equation~\eqref{eq2} yields a square system of $L = 2N +1$ coupled equations. After obtaining the coefficients $\alpha_{n,n'}$, we directly compute the absorption cross section of the individual scatterer, as described in the Supporting Information.
To evaluate the response of the array, we employ the coupled dipole model \cite{KK20,ama81}. Specifically, exploiting the periodicity of the array, we express the dipole induced in the scatterer located at position $\mathbf{R}_i$ as
\begin{equation}
\mathbf{p}_{n} {\rm e}^{{\rm i} \mathbf{k}_{\parallel} \cdot \mathbf{R}_i } = \sum_{n'=-N}^{N} \alpha_{n,n'} \left[ \mathbf{E}_{n'} {\rm e}^{{\rm i} \mathbf{k}_{\parallel} \cdot \mathbf{R}_i } + \sum_{j\neq i} \mathbf{G}(\mathbf{R}_i-\mathbf{R}_j,\omega_{n'}) \mathbf{p}_{n'} {\rm e}^{{\rm i} \mathbf{k}_{\parallel} \cdot \mathbf{R}_j }\right]. \nonumber
\end{equation}
where $\mathbf{G}(\mathbf{R},\omega) = (k^2\mathbf{I}_{3\times3}+\nabla\otimes\nabla)\exp({\rm i}k|\mathbf{R}|)/|\mathbf{R}|$ is the vacuum Green tensor, $\mathbf{I}_{3\times3}$ is the identity matrix, and $k=\omega/c$. The solution of this equation is given by
\begin{equation}
\mathbf{p}_{n}  = \sum_{n'=-N}^{N}\left(\left[\overline{\overline{\alpha}}^{\,-1}\otimes\mathbf{I}_{3\times3} - \mathbf{I}_{L\times L}\otimes\bm{\mathcal{G}}(\mathbf{k}_{\parallel}, \omega_{n'})\right]^{-1}\right)_{n,n'} \mathbf{E}_{n'}.\nonumber
\end{equation}
In this expression, $\overline{\overline{\alpha}}$ is the $L\times L$ matrix whose elements are $\alpha_{n,n'}$, while $\bm{\mathcal{G}}(\mathbf{k}_{\parallel},\omega_{n'})  = \sum_{j\neq 0}  \mathbf{G}(\mathbf{R}_j,\omega_{n'})\exp(-{\rm i}\mathbf{k}_{\parallel}\cdot \mathbf{R}_j) $ is the lattice sum \cite{KK20,ama81}. Once the induced dipoles are known, we obtain the array absorbance by integrating the Poynting vector, as detailed in the Supporting Information. 

Prior to analyzing the optical response of time-modulated arrays, we establish a reference by examining the absorbance of three time-invariant arrays (i.e, $\Delta = 0$), as shown in Figure~\ref{fig1}(b). We assume that the arrays are excited by a plane-wave electric field that propagates parallel to the $z$ axis and is polarized along the $x$ axis. The red solid curve corresponds to an array with period $a = 1.1\lambda_{\rm r}$, where $\lambda_{\rm r} = 2\pi c/\omega_{\rm r}$, and nonradiative damping rate $\gamma = \omega_{\rm r}/40$. Similarly, the green and blue curves represent arrays with $a = 1.1\lambda_{\rm r}$, $\gamma = \omega_{\rm r}/400$, and $a = 1.3\lambda_{\rm r}$, $\gamma = \omega_{\rm r}/40$, respectively. In all cases, and throughout the remainder of this work, we fix the radiative damping rate to $\kappa = 0.15\omega_{\rm r}$. The spectra of the two arrays with $\gamma = \omega_{\rm r}/40$ exhibit a pronounced lattice resonance with a peak absorbance of $\mathcal{A} \approx 0.5$, corresponding to the theoretical maximum for a two-dimensional array of electric dipoles \cite{ama91}. However, the more collective nature of the lattice resonance supported by the array with the larger period \cite{ama68,ama83} yields a quality factor (defined as the ratio of the resonance wavelength to its linewidth) of $Q \approx 900$, which is an order of magnitude higher than that of the other array. The array with reduced nonradiative losses ($\gamma = \omega_{\rm r}/400$, green curve) also supports a lattice resonance; however, the decrease in $\gamma$ expectedly results in a significantly lower peak absorbance, $\mathcal{A} \approx 0.14$, and a higher quality factor, $Q \approx147$, than that of the array with the same period.

As shown in the inset, the three analyzed arrays also exhibit a secondary peak near $\omega_{\rm r}$. This feature, which is considerably broader and weaker than the lattice resonance, corresponds to the intrinsic resonance of the individual scatterers, modified by their mutual interactions \cite{ama81,ama83}. This interpretation is supported by comparing these absorbance spectra with the absorption cross section of an individual scatterer,  $\sigma_{\rm abs}$, plotted using an orange dashed curve, normalized to $a^2$ (assuming $a = 1.1\lambda_{\rm r}$).

The results presented in Figure~\ref{fig1}(b) underscore the remarkable properties of the lattice resonances supported by periodic arrays of scatterers, which yield optical responses that are both stronger and spectrally narrower than those of the individual constituents. In the following, we investigate how temporal modulation influences the properties of lattice resonances and how these collective modes can be exploited to enhance the characteristic phenomena associated with time-varying photonic systems. To this end, we begin by analyzing the optical response of a time-modulated individual scatterer. As detailed in the Supporting Information, the simplicity of our dipole-based formalism enables the determination of the complex eigenfrequencies governing its optical response. In essence, the system supports two stable modes and one unstable mode. The latter is associated with the runaway solution arising from the third-order time derivative of the dipole moment in Equation~\eqref{eq1}, which accounts for radiative losses \cite{J1975,AYS11}, and is located far from the real axis, thereby contributing negligibly to our results. Each of these modes has as many replicas as harmonics in the Floquet expansion, shifted by $\Omega$ in their real part. Below, we concentrate on the two stable modes within the first temporal Brillouin zone.

\begin{figure}[h!]
\begin{center}
\includegraphics[width=0.5\textwidth]{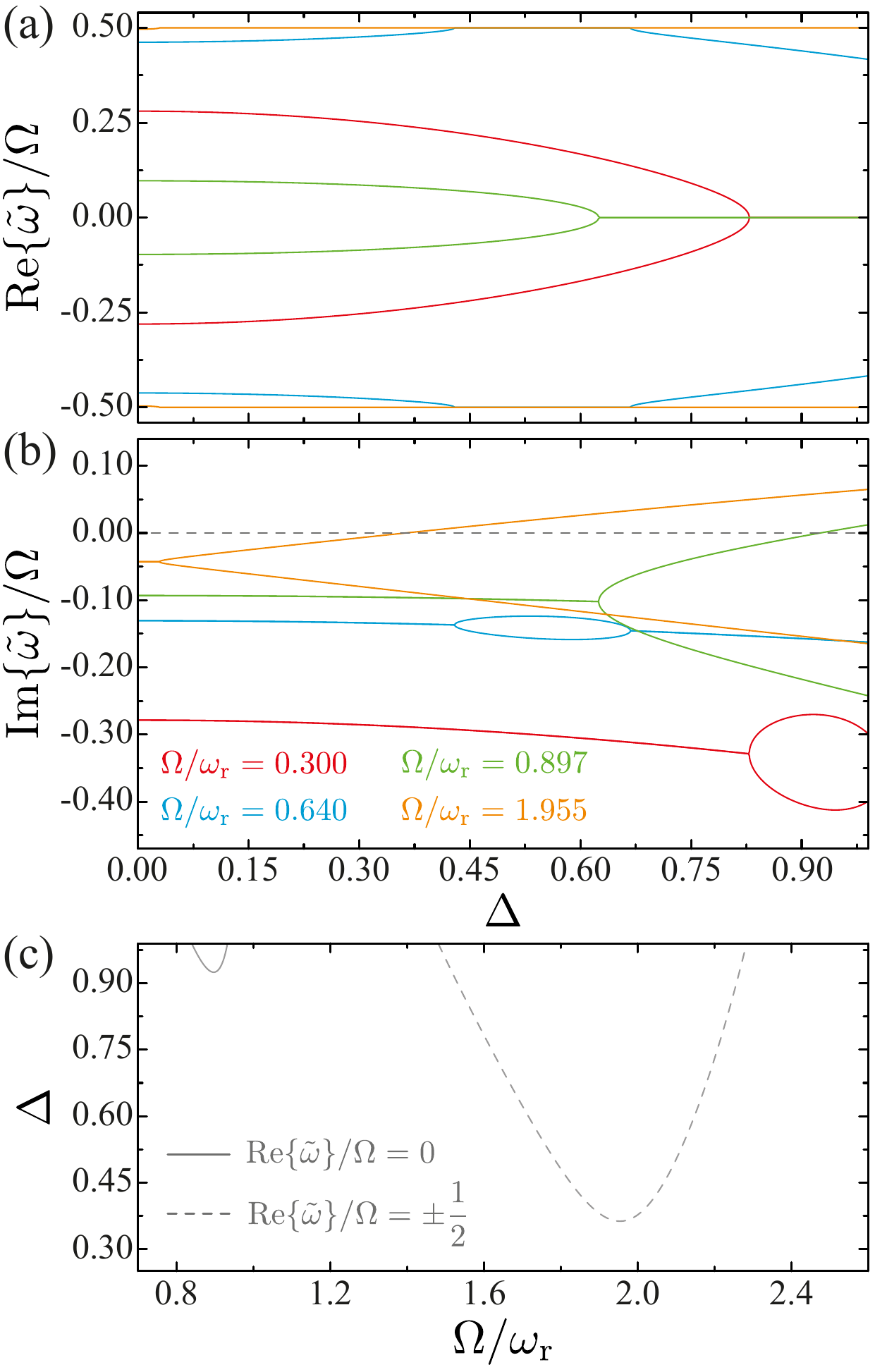}
\caption{ Real (a) and imaginary (b) parts of the complex eigenfrequencies $\tilde{\omega}$ of a time-modulated individual scatterer, plotted as a function of the modulation amplitude $\Delta$. Each colored curve corresponds to a different modulation frequency $\Omega$, as indicated in the legend. (c) Values of $\Delta$ and $\Omega$ for which one eigenfrequency crosses the real axis, with ${\rm Re}\{\tilde{\omega}\}/\Omega = 0$ (solid curve) or ${\rm Re}\{\tilde{\omega}\}/\Omega = \pm 1/2$ (dashed curve). In all cases, we assume $\gamma = \omega_{\rm r}/40$.} \label{fig2}
\end{center}
\end{figure}

Figure~\ref{fig2}(a) and Figure~\ref{fig2}(b) show, respectively, the real and imaginary parts of the eigenfrequencies $\tilde{\omega}$ of an individual scatterer as a function of the modulation amplitude $\Delta$. We assume $\gamma = \omega_{\rm r}/40$ and consider different modulation frequencies as indicated in the legend. Starting from the smallest modulation frequency, $\Omega/\omega_{\rm r} = 0.300$ (red curve), we observe the appearance of an exceptional point at approximately $\Delta = 0.829$, where the two branches of the real part of the eigenfrequencies coalesce at the center of the first temporal Brillouin zone, ${\rm Re}\{\tilde{\omega}\}/\Omega = 0$, while the corresponding imaginary part, ${\rm Im}\{\tilde{\omega}\}$, splits into two distinct branches. When the modulation frequency increases to $\Omega/\omega_{\rm r} = 0.640$ (blue curve), there is a finite range of $\Delta$ for which ${\rm Re}\{\tilde{\omega}\}$ coalesces at the edge of the first temporal Brillouin zone, ${\rm Re}\{\tilde{\omega}\}/\Omega = \pm 1/2$. In both cases, although one branch of ${\rm Im}\{\tilde{\omega}\}$ shifts closer to the real axis, its value remains negative, meaning that the amplitude of the associated eigenmode decays in time. 

The situation changes drastically when the modulation frequency further increases to $\Omega/\omega_{\rm r} = 0.897$ (green curve). Here, after the exceptional point, where the real parts coalesce again at the center of the first temporal Brillouin zone, one branch of ${\rm Im}\{\tilde{\omega}\}$ crosses the real axis at approximately $\Delta = 0.925$. This crossing signals the appearance of an amplification regime, in which the amplitude of the associated eigenmode grows without bound under external excitation. A similar behavior occurs for the highest modulation frequency considered, $\Omega/\omega_{\rm r} = 1.955$ (orange curve). In this case, the exceptional point appears approximately at $\Delta = 0.029$, with ${\rm Re}\{\tilde{\omega}\}$ coalescing at the edge of the first temporal Brillouin zone, and one branch of ${\rm Im}\{\tilde{\omega}\}$ crossing the real axis at $\Delta = 0.363$, again marking the occurrence of amplification.

To complete our analysis, Figure~\ref{fig2}(c) shows the pairs of $\Delta$ and $\Omega$ for which one eigenfrequency crosses the real axis. The solid curve corresponds to the cases with ${\rm Re}\{\tilde{\omega}\}/\Omega = 0$ (center of the temporal Brillouin zone), while the dashed curve corresponds to those with ${\rm Re}\{\tilde{\omega}\}/\Omega = \pm 1/2$ (edge of the temporal Brillouin zone). We emphasize that no additional solutions appear beyond these two sets for larger values of $\Omega/\omega_r$. 
Crossings indicated by the solid curve occur at modulation frequencies near $\omega_{\rm r}$ and, when the scatterer is excited, lead to processes with a strong contribution from the $n=-2$ harmonic. In contrast, those marked by the dashed curve, which appear closer to $\omega_{\rm r}$, result in processes where the $n = -1$ harmonic plays a significant role upon excitation.
This dependence on a lower-order harmonic explains why the cases marked by the dashed curve require a significantly smaller modulation amplitude than those corresponding to the solid curve, as clearly illustrated in Figure~\ref{fig2}(c).

\begin{figure}[h!]
\begin{center}
\includegraphics[width=\textwidth]{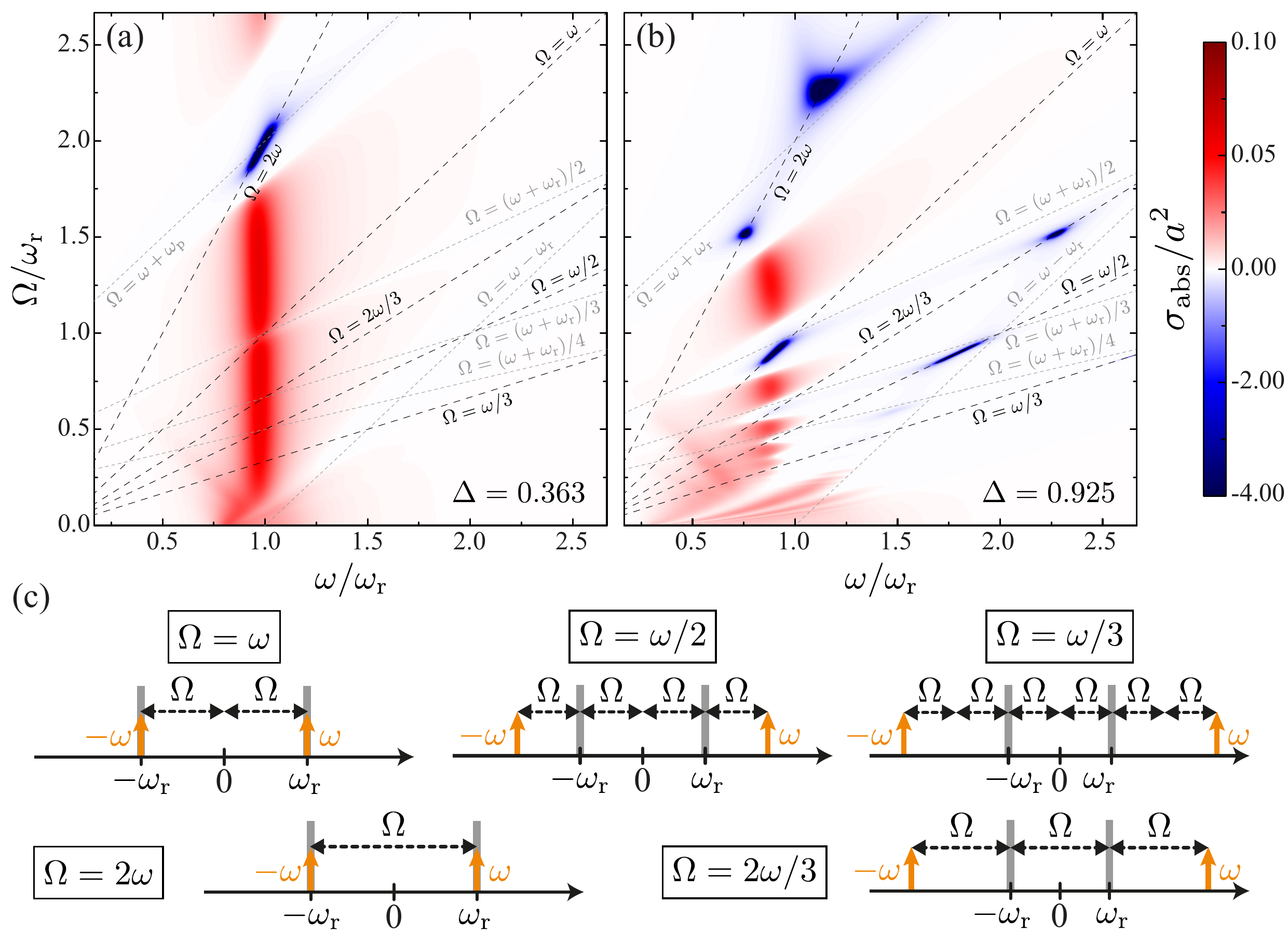}
\caption{Absorption cross section of a time-modulated individual scatterer as a function of the excitation and  modulation frequencies. We consider two modulation amplitudes: $\Delta = 0.363$ (a) and $\Delta = 0.925$ (b), corresponding to the two minima in Figure~\ref{fig2}(c). The black dashed lines indicate the conditions $\Omega = \omega / n$ and $\Omega = 2\omega / (2n\pm1)$, while the dotted gray lines correspond to $\Omega = (\omega\pm\omega_{\rm r})/n$. In both panels, $\sigma_{\rm abs}$ is normalized to $a^2$ (with $a = 1.1\lambda_{\rm r}$), and we set $\gamma = \omega_{\rm r}/40$. 
(c) Schematic illustration of the amplification process for the singularities discussed in the text. The upper row shows cases arising from the condition $\Omega=\omega/n$, while the lower row corresponds to those associated with $\Omega = 2\omega/(2n \pm 1)$. Each schematic illustrates the relationship between $\omega$ and $\Omega$, as well as their approximate connection to $\omega_{\rm r}$ at the onset of the corresponding singularity.
} \label{fig3}
\end{center}
\end{figure}

The eigenfrequency analysis presented in Figure~\ref{fig2} identifies the modulation conditions under which the individual scatterer can reach an amplification regime. Specifically, each pair of $\Delta$ and $\Omega$ in Figure~\ref{fig2}(c) determines a set of external excitation frequencies for which amplification can occur. These are given by $\omega = \mathrm{Re}\{\tilde{\omega}\} + n\Omega$, which leads to $\Omega = \omega/n$ for the condition $\mathrm{Re}\{\tilde{\omega}\}/\Omega = 0$, and $\Omega = 2\omega/(2n \pm 1)$ for $\mathrm{Re}\{\tilde{\omega}\}/\Omega = \pm 1/2$. To illustrate this behavior, we compute the absorption cross section of a time-modulated individual scatterer as a function of the excitation and modulation frequencies. Figure~\ref{fig3}(a) and Figure~\ref{fig3}(b) present the results for $\Delta = 0.363$ and $\Delta = 0.925$, respectively. These two modulation amplitudes correspond to the minima in Figure~\ref{fig2}(c), while additional cases are provided in Figure~\ref{figS1} of the Supporting Information. To facilitate comparison with the periodic arrays discussed later, we normalize the results by $a^2$, with $a = 1.1\lambda_{\rm r}$. 

Examining Figure~\ref{fig3}(a), we notice that, for small modulation frequencies, the spectrum is dominated by the resonance of the individual scatterer appearing at $\omega_{\rm r}$. However, for higher modulation frequencies a singularity emerges in the spectrum where the absorption goes to negative infinite (the color scale is saturated to improve visibility). This singularity, which is the signature of amplification, corresponds to the modulation amplitude and frequency at the minima of the gray dashed curve in Figure~\ref{fig2}(c). It appears approximately at the intersection of the conditions $\Omega = 2\omega$ and $\Omega = \omega + \omega_{\rm r}$. Indeed, identifying the crossings of $\Omega = \omega / n$ and $\Omega = 2\omega / (2n \pm 1)$ (black dashed lines) with $\Omega = (\omega \pm \omega_{\rm r})/n$ (gray dotted lines), which signals the resonant response of the system, provides a straightforward approach to approximately pinpoint the onset of all singularities associated with amplification.

Increasing the modulation amplitude to $\Delta = 0.925$ results in a significantly richer behavior. The singularity identified for $\Delta=0.363$ splits into two distinct ones that move along the $\Omega = 2\omega$ line. Furthermore, a replica of the singularity moving toward smaller $\Omega$ appears along the $\Omega=2\omega/3$ line (i.e., blueshifted by exactly $\Omega$), within the region bounded by $\Omega = (\omega+\omega_{\rm r})/2$ and $\Omega = \omega - \omega_{\rm r}$. All of these singularities correspond to the condition represented by the dashed curve in Figure~\ref{fig2}(c), for which the eigenfrequencies satisfy ${\rm Re}\{\tilde{\omega}\}/\Omega = \pm 1/2$ leading to $\Omega = 2\omega/(2n\pm1)$.  Additional singularities emerge at smaller modulation frequencies, associated with the condition indicated by the solid curve in Figure~\ref{fig2}(c), where the eigenfrequencies satisfy ${\rm Re}\{\tilde{\omega}\}/\Omega = 0$ and consequently $\Omega = \omega/n$. This is the case of the  singularity that appears close to the crossing of $\Omega = \omega$ with $\Omega = (\omega + \omega_{\rm r})/2$, together with its corresponding replicas. The latter appear, respectively, near the intersections of $\Omega = \omega/2$ with $\Omega = \omega - \omega_{\rm r}$ and $\Omega = (\omega+\omega_{\rm r})/3$, and of $\Omega = \omega/3$ with $\Omega = (\omega+\omega_{\rm r})/4$. 

To complement this analysis and facilitate its understanding, Figure~\ref{fig3}(c) provides a schematic illustration of the process leading to amplification for each of the singularities discussed above. The schematics in the upper row correspond to the singularities arising from the condition $\Omega = \omega/n$, while those in the lower row correspond to those associated with the condition $\Omega = 2\omega/(2n \pm 1)$. In all cases, the approximate relation of $\omega$ and $\Omega$ with $\omega_{\rm r}$ illustrated by each of the schematics corresponds to the onset of the corresponding singularity. 

The analysis of Figure~\ref{fig3} highlights the rich optical response of a time-modulated individual scatterer, demonstrating its ability to achieve amplification across a range of modulation amplitudes and frequencies. However, these results also indicate that modulation amplitudes exceeding $0.3$ are required to achieve that regime. This is a direct consequence of the relatively weak and broad resonance of the isolated scatterer, as shown in the inset of Figure~\ref{fig1}. It is therefore reasonable to expect that the enhanced light-matter interaction arising from the collective nature of lattice resonances, combined with their much narrower lineshapes, enables amplification at substantially smaller modulation amplitudes.

\begin{figure}[h!]
\begin{center}
\includegraphics[width=\textwidth]{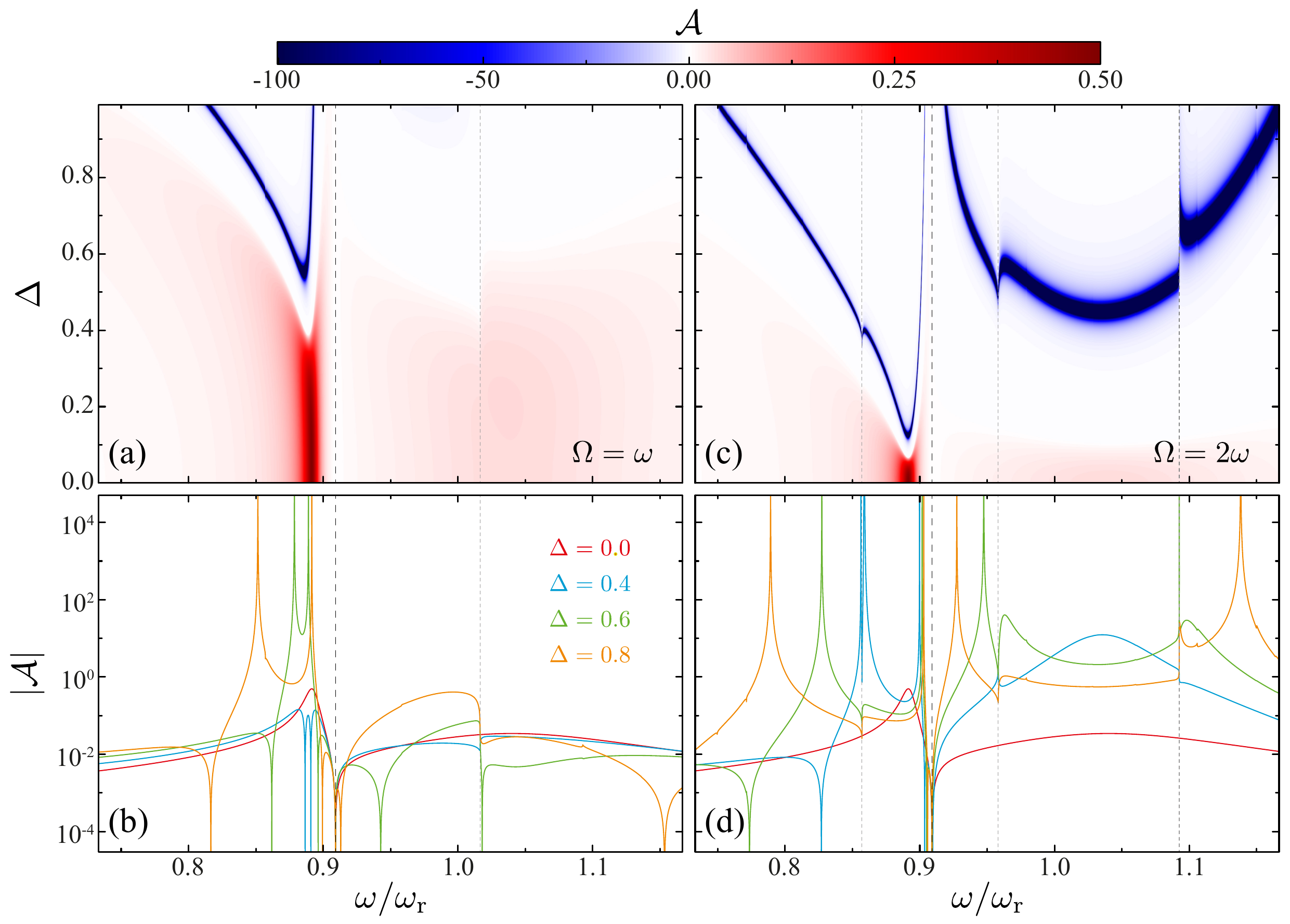}
\caption{(a,c) Absorbance of an array of time-modulated scatterers as a function of the excitation frequency and modulation amplitude. (b,d) Spectra of the absolute value of the absorbance for different values of $\Delta$, as indicated in the legend. Panels (a) and (b) show results for $\Omega = \omega$, whereas in panels (c) and (d), $\Omega = 2\omega$. In all cases, $a = 1.1\lambda_{\rm r}$ and $\gamma = \omega_{\rm r}/40$. The black dashed lines indicate the position of the first Rayleigh anomaly of the array, while the gray dotted lines correspond to replicas of higher-order Rayleigh anomalies.} \label{fig4}
\end{center}
\end{figure}

We explore this possibility in Figure~\ref{fig4}. Panel (a) displays the absorbance of an array of time-modulated scatterers with period $a = 1.1\lambda_{\rm r}$ and nonradiative damping rate of $\gamma = \omega_{\rm r}/40$. We plot the absorbance as a function of the excitation frequency and the modulation amplitude, using a modulation frequency that satisfies $\Omega = \omega$. In the absence of time modulation (i.e., $\Delta = 0$), the spectrum displays a strong absorption feature corresponding to the lattice resonance, which appear near the first Rayleigh anomaly (black dashed line) and approximately reaches a peak of $0.5$. This absorbance spectrum coincides with the one displayed by the red curve in Figure~\ref{fig1}, which we reproduce in Figure~\ref{fig4}(b) using the same color. 

Increasing the modulation amplitude to $\Delta = 0.4$ markedly reduces the absorption peak, yielding small negative values near the lattice resonance frequency (see the blue curve in Figure~\ref{fig4}(b)). For larger $\Delta$, a singularity emerges in the spectrum where the absorbance diverges toward negative infinity (the color scale is saturated for clarity). This singularity, which marks the onset of amplification, splits into two branches as the modulation amplitude continues to grow. One branch shifts towards lower frequencies, similar to the behavior observed in the eigenfrequency analysis of the time-modulated individual scatterer, indicated by the solid gray curve in Figure~\ref{fig2}(c). The other branch, however, remains practically fixed in frequency (cf. the green and orange curves in Figure~\ref{fig4}(b)), a behavior that is a direct consequence of the properties of lattice resonances. In particular, the first-order lattice resonance supported by a periodic array is always located below the frequency of the first Rayleigh anomaly, $\omega_{\rm RA} = 2\pi c/ a$, indicated here by the black dashed line. At the Rayleigh anomaly, the lattice sum diverges, causing the array to decouple from the exciting field and effectively become transparent \cite{ama68,ama81,ama83}. We also notice a replica of a higher-order Rayleigh anomaly, indicated by the gray dotted line, where the absorbance spectrum does not vanish but is strongly modified. At these replicas, which appear at frequencies $\omega = -n\Omega + \omega_{\rm RA} \sqrt{i^2+j^2}$ (with $i$ and $j$ being integers and $|n|>0$), only higher harmonics are affected, allowing the array to remain excitable by the external field. 

As anticipated, the collective nature of the lattice resonance enables a substantial reduction in the modulation amplitude required to achieve amplification: from $\Delta = 0.925$ for the individual scatterer to $\Delta = 0.551$ for the array. A similar behavior is observed for $\Omega = 2\omega$. In this case, analyzed in Figures~\ref{fig4}(c) and (d), the modulation amplitude at which the onset of amplification driven by the lattice resonance appears is reduced from $\Delta = 0.363$ to approximately $\Delta = 0.125$. As before, we observe two branches, one of which is limited by the presence of the first Rayleigh anomaly. We also identify a second singularity associated with the localized resonance of individual scatterers, which appears beyond the first Rayleigh anomaly and requires significantly larger values of $\Delta$. The shape and location of this singularity are highly consistent with the eigenfrequency analysis indicated by the dashed curve in Figure~\ref{fig2}(c), supporting our interpretation of its origin. However, its characteristics are significantly affected by the presence of several replicas of higher-order Rayleigh anomalies marked with gray dotted lines. As shown in Figure~\ref{fig4}(d), the entire absorbance spectrum exhibits pronounced variations around the frequency of each of these replicas. An interesting consequence of this effect is that some of the amplification peaks, such as the one appearing near $\omega \approx 1.09\omega_{\rm r}$ for $\Delta = 0.6$, exhibit an extremely narrow lineshape.

\begin{figure}[h]
\begin{center}
\includegraphics[width=0.5\textwidth]{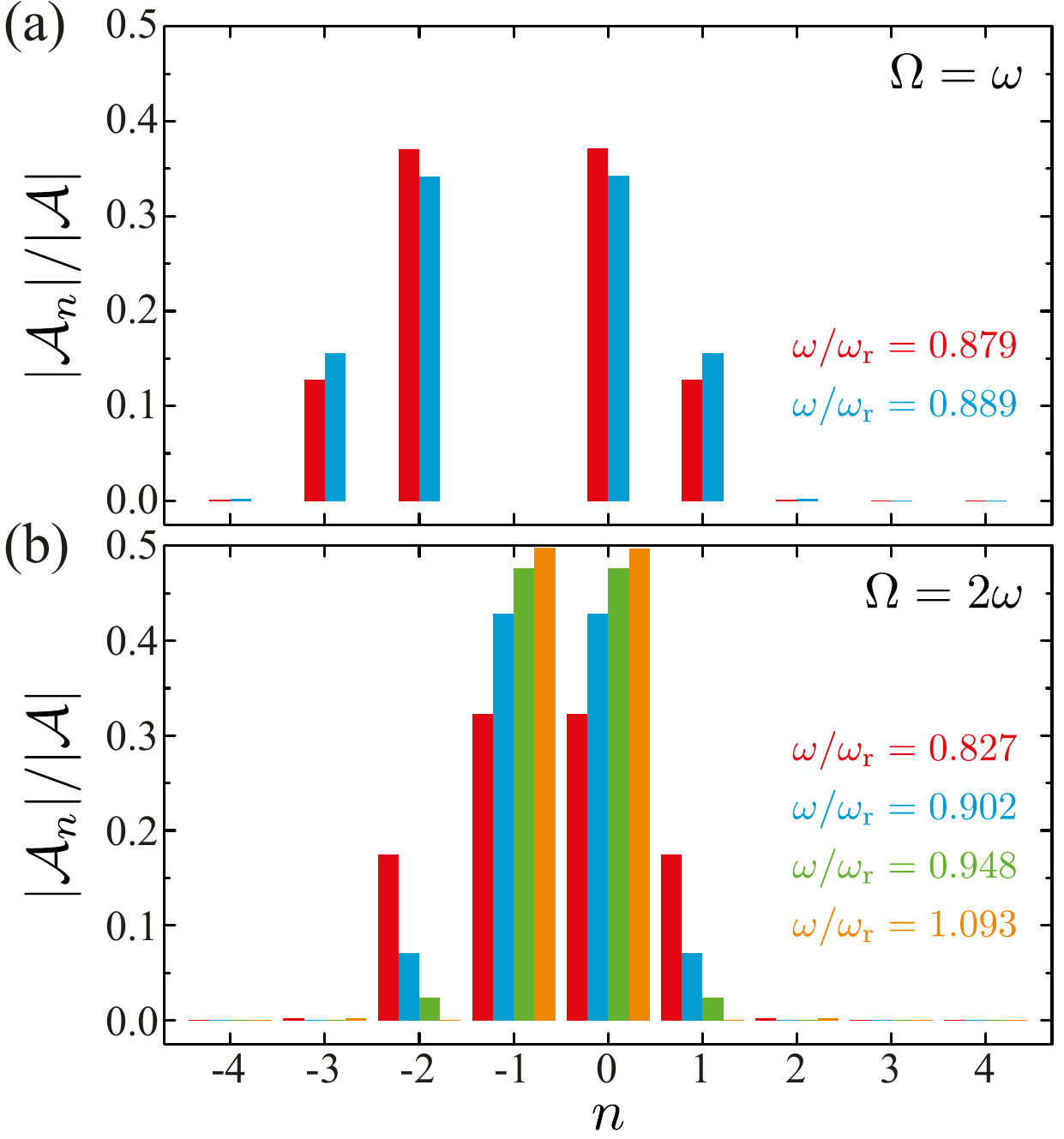}
\caption{Relative contribution of individual harmonics to the absorbance of the time-modulated array. Panels (a) and (b) correspond to $\Omega = \omega$ and $\Omega = 2\omega$, respectively. In both panels each color represents a different excitation frequency, as indicated in the legend, corresponding to the singularity peaks of the results for $\Delta = 0.6$ (green curve) shown in Figures~\ref{fig4}(b) and (d). In all cases, $a = 1.1\lambda_{\rm r}$ and $\gamma = \omega_{\rm r}/40$.} \label{fig5}
\end{center}
\end{figure}

The results in Figure~\ref{fig4} show that achieving amplification under the condition $\Omega = \omega$ requires a larger modulation amplitude than under $\Omega = 2\omega$. This difference arises from the specific harmonics involved in each case. To gain a deeper understanding of their behavior, Figure~\ref{fig5} presents the relative contributions of the individual harmonics to the absorbance of the time-modulated array (see the Supporting Information for details). Figure~\ref{fig5}(a) focuses on the condition $\Omega = \omega$, with each set of colored bars corresponding to a frequency at which a singularity appears in the absorbance spectrum (green curve in Figure~\ref{fig4}(b)). These results show that, under this condition, the $n = 0$ and $n = -2$ harmonics dominate, while the $n = -3$ and $n = 1$ harmonics contribute between one-half and one-third as much. The remaining harmonics have a negligible impact. In particular, the $n=-1$ harmonic is identically zero since no radiation is produced in the static regime. The situation differs for $\Omega = 2\omega$. In this case, we analyze four different frequencies corresponding to the singularities of the green curve in Figure~\ref{fig4}(d). As shown in Figure~\ref{fig5}(b), the largest contributions come from the $n=0$ and $n=-1$ harmonics. Additional contributions from the $n=-2$ and $n=1$ harmonics are more pronounced for the singularities driven by the lattice resonance (red and blue bars) than for those associated with the localized resonance of the individual scatterers (green and orange bars). It should be noted that the distributions are symmetric with respect to $\omega = 0$ in all cases, as required by the reality condition of the absorbance. Furthermore, these results corroborate that, under the modulation conditions considered herein, the Floquet expansion may be reliably truncated at $N = 10$ without compromising accuracy.

\begin{figure}[h!]
\begin{center}
\includegraphics[width=0.5\textwidth]{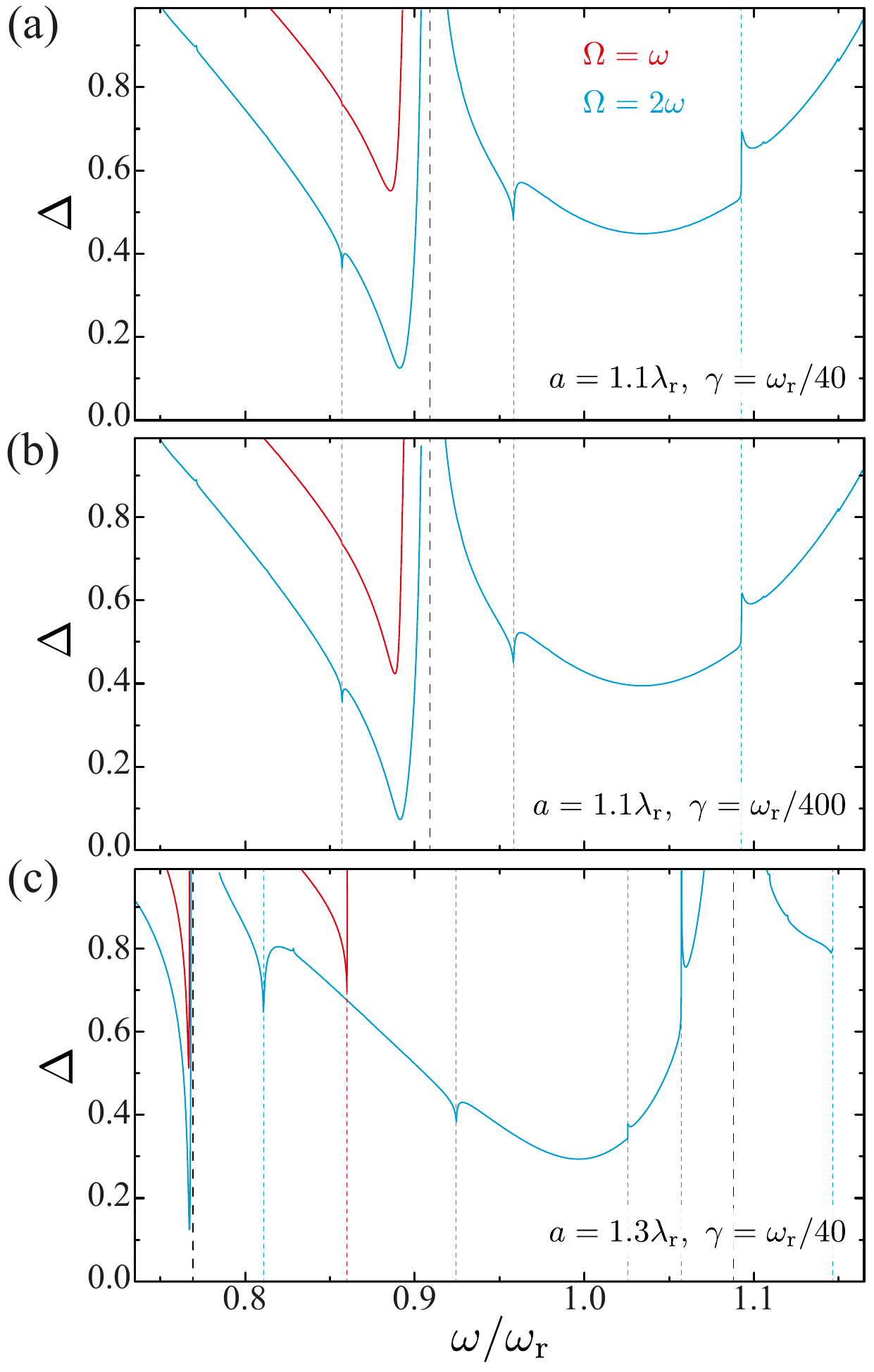}
\caption{Modulation amplitude at which the time-modulated array exhibits amplification, plotted as a function of the excitation frequency. In each panel, red and blue curves correspond, respectively, to the conditions $\Omega = \omega$ and $\Omega = 2\omega$. Each panel displays the results for different values of the period $a$ and the nonradiative damping rate $\gamma$, as indicated by the legends. The black dashed lines indicate the position of the first and second Rayleigh anomalies of the array, while the red and blue dotted lines correspond to replicas of higher-order Rayleigh anomalies.} \label{fig6}
\end{center}
\end{figure}

Thus far, our analysis has focused on a specific time-modulated array with period $a = 1.1\lambda_{\rm r}$ and nonradiative damping rate $\gamma = \omega_{\rm r}/40$. However, as indicated by the optical response of time-invariant arrays discussed in Figure~\ref{fig1}(b), both parameters exert a strong influence on the behavior of the system. To assess their impact on the amplification process, we examine three arrays in Figure~\ref{fig6}, with the same values of $a$ and $\gamma$ as in Figure~\ref{fig1}(b). In each panel, the red and blue curves indicate the values of $\Delta$ and $\omega$ for which the time-modulated array exhibits amplification under the conditions $\Omega = \omega$ and $\Omega = 2\omega$, respectively. 
The black dashed lines mark the first and second Rayleigh anomalies of the array (the second is only visible in panel (c)), while the red and blue dotted lines indicate replicas of higher-order Rayleigh anomalies.

As a reference, Figure~\ref{fig6}(a) shows the results for the array analyzed in Figures~\ref{fig4} and \ref{fig5}, while Figure~\ref{fig6}(b) corresponds to an array with the same period but a reduced nonradiative loss rate of $\gamma = \omega_{\rm r}/400$. As anticipated, decreasing $\gamma$ significantly reduces the modulation amplitude $\Delta$ required to achieve amplification. This effect occurs for both singularities driven by the lattice resonance and those linked to the localized resonance of the individual scatterers. In particular, the onset for the amplification driven by the lattice resonance occurs approximately at $\Delta = 0.423$ for $\Omega = \omega$, and at $\Delta = 0.073$ for $\Omega = 2\omega$. These results confirm that nonradiative losses are one of the most relevant limiting factors that must be optimized to achieve amplification.

Figure~\ref{fig6}(c) examines an array with the same $\gamma$ as in Figure~\ref{fig6}(a) but an increased period of $a = 1.3\lambda_{\rm r}$. As shown in Figure~\ref{fig1}(b), increasing the period leads to a lattice resonance with a significantly higher quality factor due to its more collective nature \cite{ama68,ama81,ama83}. The main consequence of this increase is that the singularity driven by the lattice resonance spans a much narrower frequency range for both $\Omega = \omega$ and $\Omega = 2\omega$, while its onset remain similar to those in Figure~\ref{fig6}(a). An important difference is that, for $\Omega = \omega$, we now observe a second singularity whose spectral shape closely resembles that of the lattice resonance and is bounded at higher frequencies by a replica of a higher-order Rayleigh anomaly (red dotted line). For $\Omega = 2\omega$, the singularity associated with the individual scatterers exhibits a more complex structure, with up to five replicas of higher-order Rayleigh anomalies altering its profile. In this case, the second-order Rayleigh anomaly appears at $\omega = \sqrt{2}\,\omega_{\rm RA}$, resulting in a range of excitation frequencies where amplification does not occur. The results in Figure~\ref{fig6} highlight the critical role of the nonradiative losses and the array period in determining the onset and spectral characteristics of the amplification process.

\section{Conclusions}

In summary, we have studied the lattice resonances supported by a periodic array of time-modulated scatterers and identified the conditions that enable  amplification. To this end, we have developed a simple yet powerful framework that combines the dipolar approximation with time-Floquet theory. In this approach, which fully accounts for dispersion as well as radiative and nonradiative losses, each scatterer is modeled as a harmonic oscillator with periodically varying optical properties, and the response of the array is described using a coupled dipole model. As a reference, we have first characterized the optical response of an individual scatterer. By analyzing the evolution of the complex eigenfrequencies of its optical modes, we have determined the modulation amplitudes and frequencies at which the imaginary part of one eigenfrequency vanishes, signaling the existence of amplification. We have confirmed this prediction by directly computing the absorption cross section of the time-modulated scatterer. 

Building on this foundation, we have extended our study to periodic arrays of time-modulated scatterers. We have computed the absorbance of the array and characterized the modulation conditions that lead to amplification. In doing so, we have found that the Rayleigh anomalies of the array (frequencies at which one diffraction order becomes grazing), as well as their replicas induced by the time modulation, have a strong effect on the spectral characteristics of the amplification. Furthermore, we have showed that the strong light-matter coupling and long lifetimes inherent to lattice resonances substantially lower the modulation strength required for amplification compared to isolated scatterers. The spectral characteristics of the amplification provided by the array depend strongly not only on its material properties but also on its period, which offers a practical degree of freedom to engineer systems that achieve amplification within a desired spectral region. The simplicity of our theoretical approach, combined with the use of normalized parameters, makes our results broadly applicable to a wide range of physical systems operating across different regions of the electromagnetic spectrum: from metallic structures in the microwave regime to doped semiconductors in the infrared and near-infrared domains, and even plasmonic systems in the visible range.

Overall, this work provides a fundamental understanding of the role of collective optical phenomena in time-varying photonics. Beyond this conceptual advance, our results introduce a novel approach to achieving strong amplification without the need for any gain material, relying instead on the interplay between lattice resonances and temporal modulation. We anticipate that these findings will inspire new research exploiting the unique properties of periodic arrays of time-modulated scatterers to develop advanced active devices with nonreciprocal and nonlinear functionalities.

\begin{acknowledgement}
This work has been supported by Grant No.~PID2022-137569NB-C42 funded by MICIU/AEI/ 10.13039/501100011033 and FEDER, EU. M.B.dP. acknowledges support from a Juan de la Cierva fellowship (Grant No.~JDC2023-050683-I) funded by MICIU/AEI/10.13039/501100011033 and the ESF+. J.R.D-R. acknowledges support from a predoctoral fellowship (Grant No.~PRE2020-095025) funded by MICIU/AEI/10.13039/501100011033 and ESF. D. M. S acknowledges support from a Ram\'on y Cajal fellowship (Grant No.~RYC2023-045265-I) funded by MICIU/AEI/10.13039/501100011033 and by the ESF+, as well as from Xunta de Galicia Regional Government (Consolidation of Competitive Research Units type C: Grant No.~ED431F 2025/22).
\end{acknowledgement}

\section{Supporting Information}

\subsubsection{Polarizability of the Time-Modulated Scatterer}

The polarizability of the individual scatterer is defined as 
\begin{equation}
\mathbf{p}(t) = \int_{-\infty}^{\infty} {\rm d}t' \alpha(t,t') \mathbf{E}(t'), \nonumber
\end{equation}
where $\alpha(t,t')=0$ for $t'>t$, ensuring causality. In the frequency domain, using the Fourier transform, we obtain
\begin{align}
\mathbf{p}(\omega) &{}= \int_{-\infty}^{\infty} {\rm d}t \,\mathbf{p}(t) {\rm e}^{{\rm i}\omega t} = \int_{-\infty}^{\infty} {\rm d}t \int_{-\infty}^{\infty} {\rm d}t'   \alpha(t,t') \mathbf{E}(t')  {\rm e}^{{\rm i}\omega t}, \nonumber \\
&{} =  \int_{-\infty}^{\infty} {\rm d}t \int_{-\infty}^{\infty} {\rm d}t'   \alpha(t,t') {\rm e}^{{\rm i}\omega t}  \int_{-\infty}^{\infty} \frac{{\rm d} \omega'}{2\pi}\mathbf{E}(\omega') {\rm e}^{-{\rm i}\omega' t'},  \nonumber \\
&{} =   \int_{-\infty}^{\infty} {\rm d}t   \int_{-\infty}^{\infty} \frac{{\rm d} \omega'}{2\pi} {\rm e}^{{\rm i}\omega t} \int_{-\infty}^{\infty} {\rm d}t'   \alpha(t,t') {\rm e}^{-{\rm i}\omega' t'}  \mathbf{E}(\omega'), \nonumber \\
&{} =    \int_{-\infty}^{\infty} \frac{{\rm d} \omega'}{2\pi}  \int_{-\infty}^{\infty} {\rm d}t\,  \alpha(t,-\omega') {\rm e}^{{\rm i}\omega t} \mathbf{E}(\omega'),   \nonumber \\
&{} =    \int_{-\infty}^{\infty} \frac{{\rm d} \omega'}{2\pi}   \alpha(\omega,-\omega') \mathbf{E}(\omega'),   \label{eqS1}
\end{align}
where
\begin{equation}
\alpha(\omega,-\omega') = \int_{-\infty}^{\infty} \int_{-\infty}^{\infty} {\rm d}t {\rm d}t' \alpha(t,t') {\rm e}^{{\rm i}\omega t}  {\rm e}^{-{\rm i}\omega' t'}. \nonumber
\end{equation}

The periodic modulation considered in this work allows us to express both the dipole and the external field using a Floquet expansion
\begin{equation}
\mathbf{p}(t) = \sum_{n=-\infty}^\infty \mathbf{p}_n {\rm e}^{-{\rm i}\omega_n t}, \nonumber
\end{equation} 
and
\begin{equation}
\mathbf{E}(t) = \sum_{n=-\infty}^\infty \mathbf{E}_n {\rm e}^{-{\rm i}\omega_n t}, \nonumber
\end{equation} 
where $\omega_n = \omega + n\Omega$, with $\omega$ denoting the excitation frequency. In the frequency domain, these become
\begin{equation}
\mathbf{p}(\omega') = \sum_{n=-\infty}^\infty \mathbf{p}_n \delta(\omega' - \omega_n), \nonumber
\end{equation} 
and 
\begin{equation}
\mathbf{E}(\omega') = \sum_{n=-\infty}^\infty \mathbf{E}_n \delta(\omega' - \omega_n). \nonumber
\end{equation} 
Substituting these expressions into Equation~\eqref{eqS1}, we find
\begin{equation}
\mathbf{p}_n = \sum_{n'=-\infty}^\infty \alpha_{n,n'} \mathbf{E}_{n'}, \nonumber
\end{equation}
where 
\begin{equation}
\alpha_{n,n'} = \alpha(\omega_n,-\omega_{n'}) = \int_{-\infty}^{\infty} \int_{-\infty}^{\infty} {\rm d}t {\rm d}t' \alpha(t,t') {\rm e}^{{\rm i}\omega_n t}  {\rm e}^{-{\rm i}\omega_{n' }t'}. \nonumber
\end{equation}

To compute the coefficients $\alpha_{n,n'}$, we start from the equation that governs the the dipole induced in the scatterer
\begin{equation}
\mathbf{\ddot{p}}(t) + \omega_{\rm r}^2 f(t) \mathbf{p}(t) + \gamma \mathbf{\dot{p}}(t) - \tau f(t) \mathbf{\dddot{p}}(t)  = \frac{3c^3}{2}  \tau f(t)  \mathbf{E}(t), \label{eqS2}
\end{equation}
where $f(t) = 1+\Delta\cos(\Omega t)$. Substituting the Floquet expansions of the dipole and the external field into Equation~\eqref{eqS2}, and noting the implicit isotropy of the system, which allows us to focus on a single Cartesian component, we obtain the following set of coupled equations:
\begin{equation}
-\omega_n^2 p_n + \omega_{\rm r}^2 \! \sum^{\infty}_{n'=-\infty} f_{n-n'} p_{n'} -{\rm i}\gamma \omega_n p_n -  {\rm i} \tau \! \sum^{\infty}_{n'=-\infty} f_{n-n'}\omega_{n'}^3 p_{n'} = \frac{3}{2} c^3 \tau \!\sum^{\infty}_{n'=-\infty} f_{n-n'} E_{n'},\nonumber
\end{equation}
where $f_n = \delta_{n,0} +\left(\delta_{n,1} +\delta_{n,-1}\right)\Delta/2$ are the Fourier coefficients of $f(t)$. Truncating the expansions at an integer $N$ yields a system of $L = 2N + 1$ coupled equations. For example, taking $N = 2$ (thus $L = 5$), we have
\begin{equation}
    \begin{pmatrix}
        A_{-2} & B_{-1} & 0 & 0 & 0 \\
        B_{-2} & A_{-1} & B_0 & 0 & 0 \\
        0 & B_{-1} & A_0 & B_{1} & 0 \\
        0 & 0 & B_0 & A_{1} & B_{2} \\
        0 & 0 & 0 & B_{1} & A_{2}
    \end{pmatrix}
    \begin{pmatrix}
        p_{-2} \\
        p_{-1} \\
        p_0 \\
        p_{1} \\
       p_{2}
    \end{pmatrix}
   = 
      \begin{pmatrix}
        F_0  & F_{-1} & 0 & 0 & 0 \\
        F_{1}& F_0  & F_{-1} & 0 & 0 \\
        0 &   F_{1} & F_0  & F_{-1}& 0 \\
        0 & 0 &   F_{1}& F_0  & F_{-1}\\
        0 & 0 & 0 & F_{1}& F_0
    \end{pmatrix}
    \begin{pmatrix}
        E_{-2} \\
        E_{-1} \\
        E_0 \\
        E_{1} \\
        E_{2}
    \end{pmatrix}.
    \label{eqS3}
\end{equation}
In these expressions, $A_n = \omega_{\rm r}^2 - \omega_n^2-{\rm i}(\gamma \omega_n + \tau \omega_n^3)$, $B_n = (\omega_{\rm r}^2 - {\rm i} \tau \omega_n^3)\Delta/2$, and $F_n = 3c^3 \tau f_n / 2$. 
Solving Equation~\eqref{eqS3}, we can write the induced dipole as
\begin{equation}
\overline{p} = \overline{\overline{\alpha}}\, \overline{E}, \nonumber
\end{equation}
where the single overbar denotes a vector of $L$ components, and a double overbar denotes a matrix of size $L\times L$. The  polarizability of the time-modulated individual scatterer is then obtained by solving:
\begin{equation}
\overline{\overline{\alpha}}=
    \begin{pmatrix}
        A_{-2} & B_{-1} & 0 & 0 & 0 \\
        B_{-2} & A_{-1} & B_0 & 0 & 0 \\
        0 & B_{-1} & A_0 & B_{1} & 0 \\
        0 & 0 & B_0 & A_{1} & B_{2} \\
        0 & 0 & 0 & B_{1} & A_{2}
    \end{pmatrix}^{-1}
      \begin{pmatrix}
        F_0  & F_{-1} & 0 & 0 & 0 \\
        F_{1}& F_0  & F_{-1} & 0 & 0 \\
        0 &   F_{1} & F_0  & F_{-1}& 0 \\
        0 & 0 &   F_{1}& F_0  & F_{-1}\\
        0 & 0 & 0 & F_{1}& F_0
    \end{pmatrix},\nonumber
\end{equation}
which is exactly equivalent to Equation~\eqref{eq2} of the main text. 

\subsubsection{Eigenmodes of the Time-Modulated Scatterer}

In principle, we can extract the complete set of eigenfrequencies (denoted as $\tilde{\omega}_i$, with $i = 1, \dots, 3L$) that characterize the response of the time-modulated scatterer by finding the roots of the determinant of $\overline{\overline{\alpha}}^{-1}$, i.e., the complex frequencies at which this matrix has at least one zero eigenvalue. However, this approach does not provide the associated eigenvectors $\tilde{\overline{p}}_i$, which contain the relative amplitudes and phases with which the $L$ dipole harmonics at $\tilde{\omega}_i + n\Omega$ jointly resonate, that is, self-sustain without an external electric field.
Moreover, because $\overline{\overline{\alpha}}^{-1}$ contains $\omega$ wrapped in terms involving $\omega_n^2$ and $\omega_n^3$, we cannot directly formulate an eigenvalue problem. To address this problem, we unwrap these terms by introducing the auxiliary variables  $j_n = -{\rm i}\omega_n p_n$ and $\phi_n = -{\rm i}\omega_n j_n = -\omega_n^2 p_n$ representing the first and second time derivatives of the dipole, respectively. By doing so, we obtain the following generalized eigenvalue problem
\begin{equation}
\begin{pmatrix}
\overline{\overline{Z}}^{\,p} & \gamma \overline{\overline{I}} & \overline{\overline{Z}}^{\,\phi}  \\
\overline{\overline{D}}  & -\overline{\overline{I}} &\overline{\overline{0}} \\
 \overline{\overline{0}}& \overline{\overline{D}} &-\overline{\overline{I}}
 \end{pmatrix}_{3L \times 3L}
\begin{pmatrix}
\overline{p} \\
 \overline{j} \\
 \overline{\phi}
 \end{pmatrix}_{3L \times 1}= \omega \begin{pmatrix}
 \overline{\overline{0}}  & \overline{\overline{0}} & \overline{\overline{W}}  \\
{\rm i} \overline{\overline{I}} & \overline{\overline{0}}  & \overline{\overline{0}}\\
 \overline{\overline{0}} & {\rm i} \overline{\overline{I}} &  \overline{\overline{0}}
 \end{pmatrix}_{3L \times 3L}
\begin{pmatrix}
\overline{p} \\
 \overline{j} \\
 \overline{\phi}
 \end{pmatrix}_{3L \times 1}.
\label{eqS4}
\end{equation}
Here, $\overline{\overline{I}}$ and $\overline{\overline{0}}$ denote the identity and zero matrices of dimensions $L\times L$, respectively. Furthermore, $\overline{\overline{D}}$, $\overline{\overline{Z}}^{\,\xi}$, and $\overline{\overline{W}}$, considering again the specific case of $N = 2$, are given by:
\begin{equation}
D =
\begin{pmatrix}
          2{\rm i} \Omega& 0 & 0 & 0 & 0 \\
        0 & {\rm i} \Omega & 0 & 0 & 0 \\
        0  & 0 & 0 & 0 & 0 \\
        0 & 0 & 0 & -{\rm i} \Omega & 0 \\
        0 & 0 & 0 & 0 & -2{\rm i} \Omega
    \end{pmatrix},
\ \
\overline{\overline{Z}}^{\,\xi}=
    \begin{pmatrix}
        X^{\xi}_{-2} & Y^{\,\xi}_{-1} & 0 & 0 & 0 \\
        Y^{\,\xi}_{-2}& X^{\xi}_{-1}  & Y^{\,\xi}_{0} & 0 & 0 \\
        0  & Y^{\,\xi}_{-1} & X^{\xi}_{0}  & Y^{\,\xi}_{1} & 0 \\
        0 & 0 & Y^{\,\xi}_{0} & X^{\xi}_{1}  & Y^{\,\xi}_{2} \\
        0 & 0 & 0 & Y^{\,\xi}_{1}& X^{\xi}_{2} 
    \end{pmatrix},
    \nonumber
\end{equation}
and 
\begin{equation}
\overline{\overline{W}}=
\begin{pmatrix}
        W_0  & W_{-1} & 0 & 0 & 0 \\
        W_{1}& W_0  & W_{-1} & 0 & 0 \\
        0 &   W_{1} & W_0  & W_{-1}& 0 \\
        0 & 0 &   W_{1}& W_0  & W_{-1}\\
        0 & 0 & 0 & W_{1}&  W_0
    \end{pmatrix}.\nonumber
    \nonumber
\end{equation}
In these expressions, $X^{p}_{n} = \omega_{\rm r}^2$, $Y^{p}_n = \omega_{\rm r}^2 \frac{\Delta}{2}$, $X^{\phi}_{n} = 1+{\rm i} \tau n \Omega$, $Y^{\phi}_n = {\rm i} \tau n\Omega \frac{\Delta}{2}$, and $W_{n} = -{\rm i}\tau f_n$. 

By solving the generalized eigenvalue problem defined by Equation~\eqref{eqS4}, we obtain $3L$ eigenfrequencies $\tilde{\omega}_i$ along with their corresponding eigenvectors $\tilde{\overline{p}}_i$. These represent three fundamental eigenmodes and their $L - 1$ replicas located outside the first temporal Brillouin zone. One of these modes corresponds to the so-called runaway mode\cite{J1975,MK1976}, a byproduct of the third-order time derivative of the dipole moment in Equation~\eqref{eqS2} accounting for radiative losses. The remaining two eigenfrequencies are those plotted and analyzed in Figure~\ref{fig2} of the main text.

\subsubsection{Modal Expansion of the Optical Response of a Time-Modulated Scatterer}

We focus, without loss of generality, on the response of an individual time-modulated scatterer to a single harmonic excitation $E_n$, such that only the $n$-th term on the right-hand side of Equation~\eqref{eqS3} is nonzero. Noting that each pole $\tilde{\omega}_i$ in our system can be associated with an oscillator, our goal is to express the $L$-harmonic optical response of the time-modulated scatterer to an input frequency $\omega$ as the sum of individual contributions from each oscillator, that is, as a linear combination of the eigenvectors $\tilde{\overline{p}}_i$:
\begin{equation}
\overline{p}(\omega) = \sum_{i=1}^{3L} \frac{\eta_i}{\omega - \tilde{\omega}_i} \tilde{\overline{p}}_i, \label{eqS5}
\end{equation}
with weighting coefficients equal to the product of the pole strength $\eta_i$ and the dispersive factor $1/(\omega - \tilde{\omega}_i)$. We stress that $\eta_i$ is a scalar constant that depends neither on $\omega$ nor on the harmonic order $n$: for each $\omega_n$, the dispersion of the time-modulated scatterer has a semi-closed form expressed as a sum of $3L$ sheets with heights $\eta_i \tilde{p}_{i,n} /(\omega - \tilde{\omega}_i)$. 

To find $\eta_i$, we resort to the adjugate (i.e., the transpose of the cofactor matrix) of $\overline{\overline{\alpha}}^{\,-1}$, satisfying
\begin{equation}
\overline{\overline{\alpha}}(\omega) = \frac{{\rm adj}\!\left[\overline{\overline{\alpha}}^{\, -1}(\omega)\right]}{{\rm det}\!\left[\overline{\overline{\alpha}}^{\, -1}(\omega)\right]}, \nonumber
\end{equation}
where we deliberately emphasize the $\omega$ dependence. Clearly, column $n$ in $\overline{\overline{\alpha}}(\omega)$ represents the contribution to the dipole $\overline{p}(\omega)$ arising from a normalized input $E_n = 1$ and can be expanded as in Equation~\eqref{eqS5}. Noting also that
\begin{equation}
{\rm det}\!\left[\overline{\overline{\alpha}}^{\, -1}(\omega)\right] = \Theta \prod_{i=1}^{3L} (\omega - \tilde{\omega}_i), \nonumber
\end{equation}
with $\Theta$ being a constant easily retrievable by sampling, we can equate
\begin{equation}
\frac{{\rm adj}\!\left[\overline{\overline{\alpha}}^{\, -1}(\omega)\right]_{{\rm column}\, n}}{\Theta \prod_{i=1}^{3L} (\omega - \tilde{\omega}_i)}
=
\frac{\sum_{i=1}^{3L}  \eta_i \tilde{\overline{p}}_i \prod_{j \neq i} (\omega - \tilde{\omega}_j)}{\prod_{i=1}^{3L} (\omega - \tilde{\omega}_i)}. \nonumber
\end{equation}
Consequently, by probing the system at $\omega = \tilde{\omega}_i$, the contributions from $\tilde{\overline{p}}_{j \neq i}$ are cancelled out, which leads to
\begin{equation}
\eta_i \tilde{\overline{p}}_i = \frac{{\rm adj}\!\left[\overline{\overline{\alpha}}^{\, -1}(\tilde{\omega}_i)\right]_{{\rm column}\, n}}{\Theta \prod_{j \neq i} (\tilde{\omega}_i - \tilde{\omega}_j)}. \nonumber
\end{equation}
This is a scaled and dephased version of the eigenmode $\tilde{\overline{p}}_i$ from the previous section.

\subsubsection{Absorption Cross Section of a Time-Modulated Scatterer}

Once the induced dipole in the scatterer is determined, we can calculate the radiated electric field at a point $\mathbf{r}$ as
\begin{equation}
\mathbf{E}^{\rm rad}(\mathbf{r}, t) = \sum_{n=-N}^{N} \mathbf{G}(\mathbf{r}, \omega_n)\mathbf{p}_n {\rm e}^{-{\rm i}\omega_n t} + {\rm c.c.}, \nonumber
\end{equation}
where $\mathbf{G}(\mathbf{r},\omega)$ denotes the vacuum Green tensor defined in the main text, and ``c.c.'' indicates the complex conjugate. The total electric field at $\mathbf{r}$ is the superposition of the incident field and the radiated field. The incident field exciting the scatterer is 
$\mathbf{E}(\mathbf{r}, t) = \mathbf{E}\exp({\rm i}\mathbf{k} \cdot \mathbf{ r} - {\rm i}\omega t) +  {\rm c.c.} $, where $\mathbf{k}$ is the wavevector of the incoming wave and $\mathbf{E}$ its amplitude. Consequently, the total field reads $\mathbf{E}^{\rm tot}(\mathbf{r}, t) = \mathbf{E}(\mathbf{r}, t) + \mathbf{E}^{\rm rad}(\mathbf{r}, t)$.
Similarly, the total magnetic field is given by $\mathbf{H}^{\rm tot}(\mathbf{r}, t) = \mathbf{H}(\mathbf{r}, t) + \mathbf{H}^{\rm rad}(\mathbf{r}, t)$, with $\mathbf{H}(\mathbf{r}, t) = (\mathbf{k}/k)\times \mathbf{E}(\mathbf{r}, t)$ and 
\begin{equation}
\mathbf{H}^{\rm rad}(\mathbf{r}, t) = -\frac{{\rm i}}{c} \sum_{n=-N}^{N} \frac1{\omega_n} \nabla\times \mathbf{G}(\mathbf{r}, \omega_n)\mathbf{ p}_n{\rm e}^{-{\rm i}\omega_n t} + {\rm c.c.}\nonumber
\end{equation}

The flux of electromagnetic energy is determined by the Poynting vector $\mathbf{S}(\mathbf{r}, t) = c/(4\pi)\mathbf{E}^{\rm tot}(\mathbf{r}, t) \times \mathbf{H}^{\rm tot} (\mathbf{r}, t)$. To calculate the total power absorbed by the scatterer, we integrate the time-averaged Poynting vector over a spherical surface that encloses it.  By dividing the absorbed power by the intensity of the incident field, we obtain the following expression for the absorption cross section in terms of the different components of the polarizability 
\begin{equation}
\sigma_{\rm abs} = 4\pi k {\rm Im}\{\alpha_{0,0}\} - \frac{8\pi}{3c^4}\sum_{n=-N}^{N} \omega_n^4 |\alpha_{n,0}|^2.
\nonumber
\end{equation}
As expected, $\sigma_{\rm abs}$ contains a sum over harmonics that accounts for the ability of the time-modulated scatterer to radiate at frequencies different from the exciting frequency. Notice that we do not consider additional contributions that arise at a discrete set of frequencies and depend on the relative phase between the time modulation and the incident field.

\subsubsection{Absorbance of a Periodic Array of  Time-Modulated Scatterers}

We exploit the periodicity of the array to compute the electric field it radiates to the far field as a sum over the different diffraction orders\cite{ama68} 
\begin{equation}
\mathbf{ E}^{\rm rad}_n(\mathbf{r}) = \sum_{\mathbf{q}}  \mathbf{E}^{\rm rad \pm}_{n, \mathbf{q}} {\rm e}^{{\rm i} \mathbf{ k}_{n, \mathbf{q}}^{\pm} \cdot \mathbf{r} },
\nonumber
\end{equation}
where
\begin{equation}
\mathbf{ E}^{\rm rad \pm}_{n, \mathbf{ q}} = \frac{2 \pi {\rm i}}{a^2} \frac{1}{k_{z, n, \mathbf{q}}}\left[\frac{\omega_n^2}{c^2}\mathbf{p}_n - (\mathbf{k}_{n, \mathbf{q}}^{\pm}\cdot\mathbf{p}_n) \mathbf{ k}_{n, \mathbf{q}}^{\pm}\right].
\nonumber
\end{equation}
In these expressions, $a$ is the period of the square array, whereas $\mathbf{q} = q_x \mathbf{\hat{x}}+q_y \mathbf{\hat{y}}$ denotes the reciprocal lattice vectors, such that
 \begin{equation}
 \mathbf{k}_{n, \mathbf{q}}^{\pm} = (k_x + q_x)\hat{\mathbf{x}} + (k_y + q_y)\hat{\mathbf{y}} \pm k_{z, n, \mathbf{q}}\hat{\mathbf{z}}, \nonumber
 \end{equation}
 with $k_{z, n, \mathbf{q}} = \sqrt{(\omega_n / c)^2 - (k_x+q_x)^2 - (k_y+q_y)^2}$. The sums only include those reciprocal lattice vectors that lead to propagating diffraction orders, i.e., those for which $k_{z, n, \mathbf{q}}$ is real. The upper and lower signs apply to $z>0$ (above the array) and $z<0$ (below the array), respectively. We express the magnetic field in a similar form
\begin{equation}
\mathbf{ H}^{\rm rad}_n(\mathbf{r}) = \sum_{\mathbf{q}}  \mathbf{H}^{\rm rad \pm}_{n, \mathbf{q}} {\rm e}^{{\rm i} \mathbf{k}_{n, \mathbf{q}}^{\pm} \cdot \mathbf{r}},
\nonumber
\end{equation}
with
\begin{equation}
\mathbf{H}^{\rm rad \pm}_{n, \mathbf{q}} = \frac{c\mathbf{k}_{n, \mathbf{q}}^{\pm}}{\omega_n} \times \mathbf{ E}^{\rm rad \pm}_{n, \mathbf{q}}.
\nonumber
\end{equation}
We then write the total electric and magnetic fields as
\begin{align}
\mathbf{ E}^{\rm tot}(\mathbf{ r}, t) &= \mathbf{ E}{\rm e}^{{\rm i} \mathbf{ k} \cdot \mathbf{r} -{\rm i} \omega t} + \sum_{n=-N}^{N} \mathbf{ E}^{\rm rad}_n(\mathbf{r}) {\rm e}^{-{\rm i} \omega_n t} + {\rm c.c.},
\nonumber \\
\mathbf{H}^{\rm tot}(\mathbf{r}, t) &= \frac{\mathbf{k}}{k}\times\mathbf{E}{\rm e}^{{\rm i} \mathbf{k} \cdot \mathbf{ r} -{\rm i} \omega t} + \sum_{n=-N}^{N} \mathbf{ H}^{\rm rad}_n(\mathbf{ r}) {\rm e}^{-{\rm i} \omega_n t} + {\rm c.c.}
\nonumber
\end{align}
As in the case of the individual scatterer, we compute the absorbed power using the Poynting vector. To enclose the array, we place two infinite planes parallel to its surface: one above the array at positive $z$, and one below at negative $z$. By integrating the time-averaged Poynting vector over these planes, we obtain the total power absorbed by the array. Normalizing this absorbed power by the power of the incident field yields the absorbance
\begin{equation}
\mathcal{A} = \sum_{n=-N}^{N} \mathcal{A}_n, 
\nonumber
\end{equation}
where the contribution of each harmonic is given by
\begin{equation}
\mathcal{A}_n =  -2 \delta_{n,0} {\rm Re} \left\{ \frac{\mathbf{ E^{*}}}{|\mathbf{E}|^2} \cdot \mathbf{ E}^{\rm rad +}_{0,0} \right\} - \sum_\mathbf{q} \frac{c k_{z, n, \mathbf{q}}}{\omega_n } \left(\frac{|\mathbf{E}^{\rm rad +}_{n, \mathbf{q}}|^2}{|\mathbf{E}|^2} + \frac{|\mathbf{E}^{\rm rad -}_{n, \mathbf{q}}|^2}{|\mathbf{E}|^2} \right).\nonumber
\end{equation}
Once again, we do not consider additional contributions that arise at a discrete set of frequencies and depend on the relative phase between the time modulation and the incident field.

\begin{figure}[h!]
\begin{center}
\includegraphics[width=0.9\textwidth]{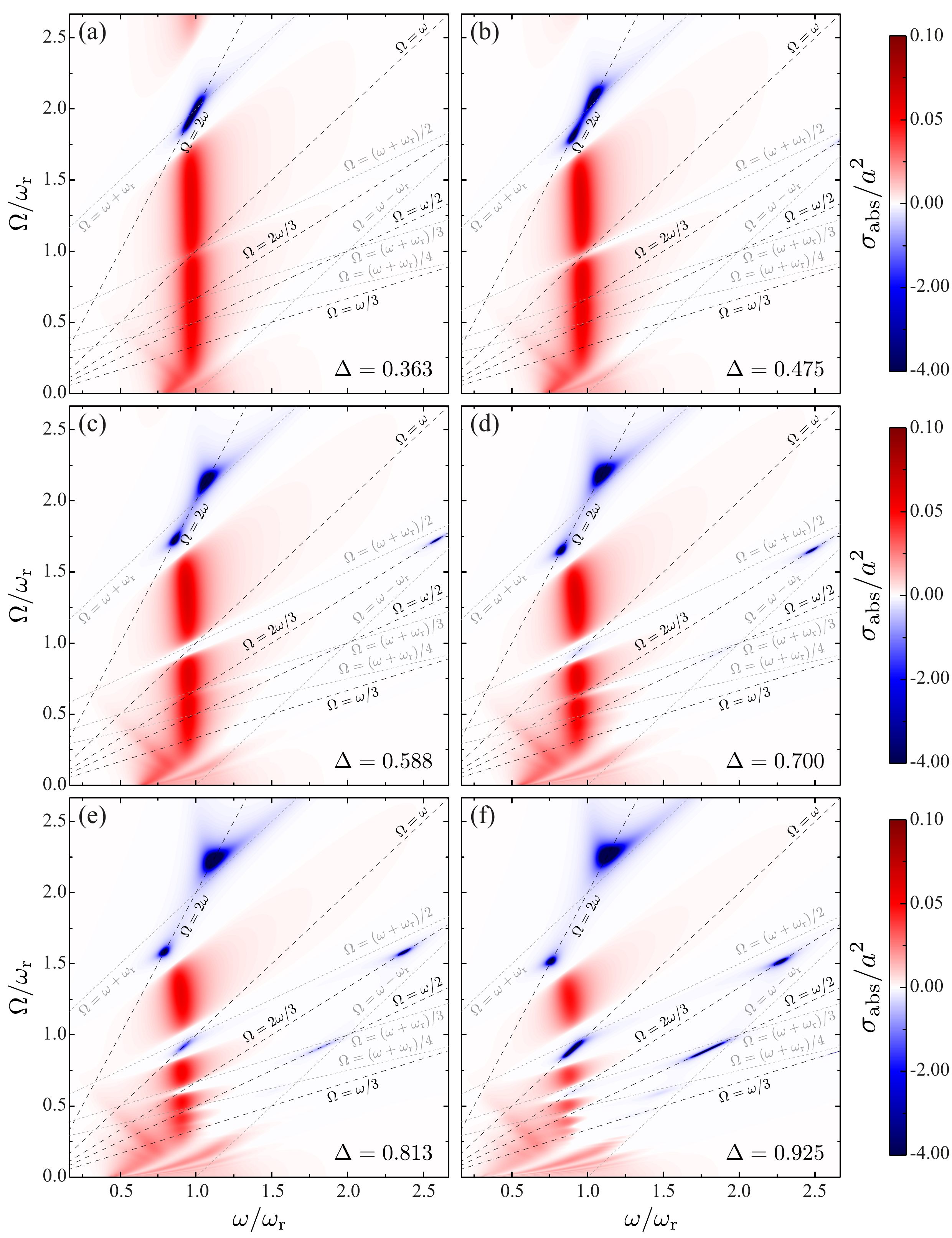}
\caption{Absorption cross section of a time-modulated individual scatterer as a function of the excitation and  modulation frequencies. We consider six modulation amplitudes: $\Delta = 0.317$ (a), $\Delta = 0.441$ (b), $\Delta = 0.565$ (c), $\Delta = 0.689$ (d), $\Delta = 0.813$ (e) and $\Delta = 0.937$ (f). The black dashed lines indicate the conditions $\Omega = \omega / n$ and $\Omega = 2\omega / (2n\pm1)$, while the dotted gray lines correspond to $\Omega = (\omega\pm\omega_{\rm r})/n$. In all panels, $\sigma_{\rm abs}$ is normalized to $a^2$ (with $a = 1.1\lambda_{\rm r}$), and we set $\gamma = \omega_{\rm r}/40$.} \label{figS1}
\end{center}
\end{figure}

\clearpage


\providecommand{\latin}[1]{#1}
\makeatletter
\providecommand{\doi}
  {\begingroup\let\do\@makeother\dospecials
  \catcode`\{=1 \catcode`\}=2 \doi@aux}
\providecommand{\doi@aux}[1]{\endgroup\texttt{#1}}
\makeatother
\providecommand*\mcitethebibliography{\thebibliography}
\csname @ifundefined\endcsname{endmcitethebibliography}
  {\let\endmcitethebibliography\endthebibliography}{}

\end{document}